\documentclass[fleqn,usenatbib,useAMS]{mnras}
 \usepackage{graphicx}
 \usepackage{amsmath}
 \usepackage{amssymb}
 \usepackage[T1]{fontenc}
 \usepackage{ae,aecompl}
 \usepackage{txfonts}
 \usepackage{enumerate}

 \title[Visible spectra of (474640)~2004~VN$_{112}$--2013~RF$_{98}$]
       {Visible spectra of (474640)~2004~VN$_\mathbf{112}$--2013~RF$_\mathbf{98}$ with OSIRIS 
        at the 10.4 m GTC: evidence for binary dissociation near aphelion among the 
        extreme trans-Neptunian objects\thanks{Based on observations made with the 
        GTC telescope, in the Spanish Observatorio del Roque de los Muchachos of the 
        Instituto de Astrof\'{\i}sica de Canarias, under Director's Discretionary 
        Time (program ID GTC2016-055).}
       }

 \author[J. de Le\'on, C. de la Fuente Marcos and R. de la Fuente Marcos]
        {J.~de~Le\'on,$^{1,2}$\thanks{E-mail: jmlc@iac.es}
         C.~de~la~Fuente~Marcos$^{3}$
         and
         R.~de~la~Fuente Marcos$^{3}$ \\
         $^1$Instituto de Astrof\'{\i}sica de Canarias, 
             C/ V\'{\i}a L\'actea s/n, E-38205 La Laguna, Tenerife, Spain \\
         $^2$Departamento de Astrof\'{\i}sica,
             Universidad de La Laguna, E-38206 La Laguna, Tenerife, Spain \\
         $^3$Universidad Complutense de Madrid,
             Ciudad Universitaria, E-28040 Madrid, Spain}
 \date{Accepted 2017 January 4. 
       Received 2016 December 23; 
       in original form 2016 October 20}
 \pubyear{2017}
 
\begin{document}
  \label{firstpage}
  \pagerange{\pageref{firstpage}--\pageref{lastpage}}
  \maketitle

  \begin{abstract}
      The existence of significant anisotropies in the distributions of the 
      directions of perihelia and orbital poles of the known extreme 
      trans-Neptunian objects (ETNOs) has been used to claim that 
      trans-Plutonian planets may exist. Among the known ETNOs, the pair 
      (474640)~2004~VN$_{112}$--2013~RF$_{98}$ stands out. Their orbital poles 
      and the directions of their perihelia and their velocities at 
      perihelion/aphelion are separated by a few degrees, but orbital 
      similarity does not necessarily imply common physical origin. In an 
      attempt to unravel their physical nature, visible spectroscopy of both 
      targets was obtained using the OSIRIS camera-spectrograph at the 10.4~m 
      Gran Telescopio Canarias (GTC). From the spectral analysis, we find that 
      474640--2013~RF$_{98}$ have similar spectral slopes (12 vs. 
      15~\%/0.1~$\mu$m), very different from Sedna's but compatible with those 
      of (148209)~2000~CR$_{105}$ and 2012~VP$_{113}$. These five ETNOs belong 
      to the group of seven linked to the Planet Nine hypothesis. A dynamical 
      pathway consistent with these findings is dissociation of a binary 
      asteroid during a close encounter with a planet and we confirm its 
      plausibility using $N$-body simulations. We thus conclude that both the 
      dynamical and spectroscopic properties of 474640--2013~RF$_{98}$ favour 
      a genetic link and their current orbits suggest that the pair was kicked 
      by a perturber near aphelion. 
  \end{abstract}


  \section{Introduction}
     The dynamical and physical properties of the extreme trans-Neptunian objects or ETNOs (semimajor axis, $a$, greater than 150 au, and 
     perihelion distance, $q$, greater than 30 au, Trujillo \& Sheppard 2014) are intriguing in many ways. Their study can help probe the 
     orbital distribution of putative planets going around the Sun between the orbit of Pluto and the Oort Cloud as well as understand the 
     formation and evolution of the Solar system as a whole. The first ETNO was found in 2000, (148209)~2000~CR$_{105}$, and its discovery 
     was soon recognised as a turning point in the study of the outer Solar system (e.g. Gladman et al. 2002; Morbidelli \& Levison 2004). 
     The current tally stands at 21 ETNOs. 

     Trujillo \& Sheppard (2014) were first in suggesting that the dynamical properties of the ETNOs could be better explained if a yet to 
     be discovered planet of several Earth masses is orbiting the Sun at hundreds of au. This interpretation was further supported by de la 
     Fuente Marcos \& de la Fuente Marcos (2014) with a Monte Carlo-based study confirming that the observed patterns in ETNO orbital 
     parameter space cannot result from selection effects and suggesting that one or more trans-Plutonian planets may exist. A plausible 
     multi-planet dynamical scenario was explored by de la Fuente Marcos, de la Fuente Marcos \& Aarseth (2015). Based on observational 
     data, and analytical and numerical work, Batygin \& Brown (2016) presented their Planet Nine hypothesis that was further developed by 
     Brown \& Batygin (2016), but questioned by Shankman et al. (2016). The orbits of seven ETNOs ---Sedna, 148209, (474640) 2004~VN$_{112}$, 
     2007~TG$_{422}$, 2010~GB$_{174}$, 2012~VP$_{113}$ and 2013~RF$_{98}$--- were used by Brown \& Batygin (2016) to predict the existence 
     of the so-called Planet Nine, most probably a trans-Plutonian super-Earth in the sub-Neptunian mass range. Out of the 21 known ETNOs, 
     only Sedna has been observed spectroscopically (see Fornasier et al. 2009). 
     
     Among the known ETNOs, the pair 474640--2013~RF$_{98}$ clearly stands out (de la Fuente Marcos \& de la Fuente Marcos 2016). The 
     directions of their perihelia (those of the vector from the Sun to the respective perihelion point) are very close (angular separation 
     of 9\fdg8), their orbital poles are even closer (4\fdg1), and consistently the directions of their velocities at perihelion/aphelion 
     are also very near each other (9\fdg5), although improved values are given in Section~3.3; in addition, they have similar aphelion 
     distances, $Q$ (589 au vs. 577 au). Assuming that the angular orbital elements of the ETNOs follow uniform distributions (i.e. they are 
     unperturbed asteroids moving in Keplerian orbits around the Sun), the probability of finding by chance two objects with such a small 
     angular separation between their directions of perihelia and, what is more important, also between their orbital poles is less than 
     0.0001, which suggests a common dynamical origin. However, a probable common dynamical origin does not imply a common physical origin. 
     In an attempt to unravel their physical nature, visible spectroscopy of the two targets was obtained on 2016 September using the OSIRIS 
     camera-spectrograph at the 10.4~m Gran Telescopio Canarias (GTC) telescope, located in La Palma (Canary Islands, Spain). Here, we 
     present and discuss the results of these observations. This Letter is organized as follows. Section~2 reviews the state of the art for 
     this pair of ETNOs. The new observations ---including spectroscopy, photometry and astrometry--- and their results are presented in 
     Section~3. The possible origin of this pair is explored in Section~4, making use of the new observational results and $N$-body 
     simulations. Conclusions are summarized in Section~5.

  \section{The pair 474640--2013~RF$_\mathbf{98}$: state of the art}
     Asteroid (474640)~2004~VN$_{112}$ was discovered on 2004 November 6 by the ESSENCE supernova survey observing with the 4~m Blanco 
     Telescope from Cerro Tololo International Observatory (CTIO) at an apparent magnitude $R$ of 22.7 (Becker et al. 2007). Its absolute 
     magnitude, $H$ = 6.4 (assuming a slope parameter, $G$ = 0.15), suggests a diameter in the range 130--300~km for an assumed albedo in 
     the range 0.25--0.05. The orbital solution (2016 August) for this object is based on 31 observations spanning a data-arc of 5113 d or 
     14 yr, from 2000 September 26 to 2014 September 26, its residual rms amounts to 0\farcs19.\footnote{Orbit available from JPL's 
     Small-Body Database and \textsc{Horizons} On-Line Ephemeris System: $a=318\pm1$ au, $e=0.8513\pm0.0005$, $i=25\fdg5748\pm0\fdg0004$, 
     $\Omega=65\fdg9990\pm0\fdg0007$ and $\omega=327\fdg121\pm0\fdg010$, referred to the epoch 2457600.5 JD TDB.} Such an object would have 
     been visible to ESSENCE for only about 2 per cent of its orbit, suggesting that the size of the population of minor bodies moving in 
     orbits similar to that of 474640 could be very significant (Becker et al. 2008). Sheppard (2010) gives a normalised spectral gradient 
     of 11$\pm$4~\%/0.1~$\mu$m for this object based on Sloan $g'$, $r'$, $i'$ optical photometry acquired in 2008. Some additional 
     photometry was obtained with the Hubble Wide Field Camera 3 (Fraser \& Brown 2012). Its optical colours are relatively neutral and this 
     was interpreted by Brown (2012) as a sign that it is not dominated by methane irradiation products.

     Asteroid 2013~RF$_{98}$ was discovered on 2013 September 12 by the Dark Energy Camera (DECam, Flaugher et al. 2015) observing from CTIO 
     for the Dark Energy Survey (DES, Abbott et al. 2005) at an apparent magnitude $z$ of 23.5 (Abbott et al. 2016). Its absolute magnitude, 
     $H$ = 8.7, suggests a diameter in the range 50--120~km. The orbital solution (2016 August) for this ETNO is rather poor as it is based 
     on 38 observations spanning a data-arc of just 56 d, from 2013 September 12 to 2013 November 7, its residual rms amounts to 
     0\farcs12.\footnote{Orbit available from \textsc{Horizons}: $a=307\pm37$ au, $e=0.882\pm0.015$, $i=29\fdg62\pm0\fdg08$, 
     $\Omega=67\fdg51\pm0\fdg13$ and $\omega=316\degr\pm6\degr$, referred to the epoch 2457600.5 JD TDB.} No other data, besides these 38 
     observations, its $H$ value, and its orbital solutions, are known about this ETNO. 

     Asteroid 474640 has been classified as an extreme detached object by Sheppard \& Trujillo (2016) and their integrations show that it is 
     a stable ETNO within the standard eight-planets-only Solar system paradigm; this result is consistent with that in Brown \& Batygin 
     (2016). In Sheppard \& Trujillo (2016), 2013~RF$_{98}$ is classified as an extreme scattered object and found to be unstable within the 
     standard paradigm over 10 Myr time-scales due to Neptune perturbations. Both ETNOs are rather unstable within some incarnations of the 
     Planet Nine hypothesis (de la Fuente Marcos, de la Fuente Marcos \& Aarseth 2016). The heliocentric orbits available in 2016 
     August$^{1,2}$ have been used to compute the values of the angular separations quoted in the previous section. In spite of the 
     limitations of the orbital solution of 2013~RF$_{98}$, the uncertainties mainly affect orbital elements other than inclination, $i$, 
     longitude of the ascending node, $\Omega$, and argument of the perihelion, $\omega$. These three orbital parameters are the only ones 
     involved in the calculation of the directions of perihelia, location of orbital poles, and directions of velocities at perihelion (de 
     la Fuente Marcos \& de la Fuente Marcos 2016). In sharp contrast, the value of $Q$ of 2013~RF$_{98}$ is affected by a significant 
     uncertainty (12 per cent).

  \section{Observations}

     \subsection{Spectroscopy}
\begin{figure}
        \centering
         \includegraphics[width=\linewidth]{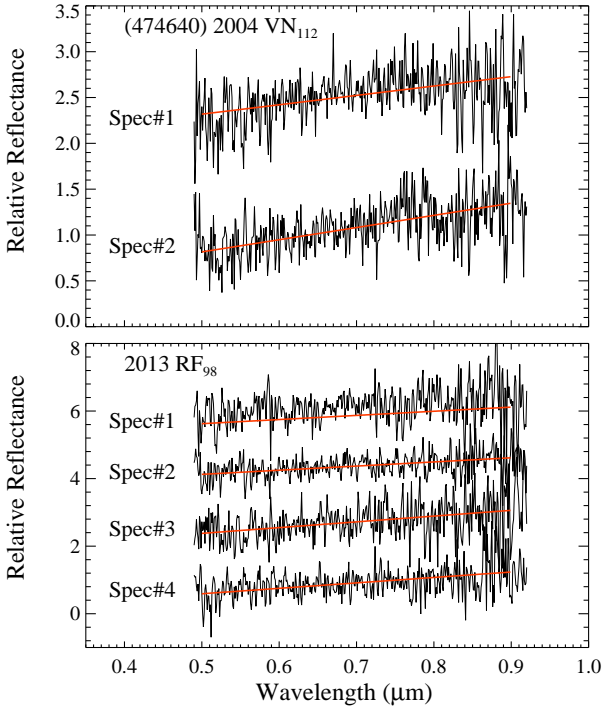}
         \caption{Individual visible spectra of (474640)~2004~VN$_{112}$ (top panel) and 2013~RF$_{98}$ (bottom panel) normalised to unity 
                  at 0.55 $\mu$m and offset vertically for clarity. Red lines correspond to the linear fitting in the range 0.5--0.9 $\mu$m 
                  to compute the spectral slope.
                 }
         \label{specs}
      \end{figure}

        Visible spectra of the two targets were obtained using OSIRIS (Cepa et al. 2000) at 10.4~m GTC. The apparent visual magnitude, $V$, 
        at the time of observation was 23.3 for (474640)~2004~VN$_{112}$ (heliocentric distance of 47.7 au and phase of 1\fdg1) and 24.4 for 
        2013~RF$_{98}$ (36.6 au, 1\fdg3). For each target, acquisition images in the Sloan $r'$ filter were obtained in separate nights in 
        order to identify reliably the object in the field of view. This procedure ended up being the most efficient to detect these dim, 
        slow-moving (apparent proper motion $<$2\arcsec/h) targets. Visible spectra were acquired using the low-resolution R300R grism 
        (resolution of 348 measured at a central wavelength of 6635~\AA\ for a 0\farcs6 slit width), that covers the wavelength range from 
        0.49 to 0.92 $\mu$m, and a 2\arcsec slit width. Two widely accepted solar analogue stars from Landolt (1992) ---SA93-101 and 
        SA115-271--- were observed using the same spectral configuration at an airmass identical to that of the targets to obtain the 
        reflectance spectra of the ETNOs. For a given ETNO, the spectrum was then divided by the corresponding spectrum of the solar analog. 
        Additional data reduction details are described by de Le\'on et al. (2016) and Morate et al. (2016). For 474640 we acquired two 
        spectra of 1800~s each, while for 2013~RF$_{98}$ we acquired four individual spectra of 1800~s each. Observational details are shown 
        in Table \ref{spec}. The resulting individual reflectance spectra, normalised to unity at 0.55~$\mu$m and offset vertically for 
        clarity, are shown in Fig.~\ref{specs}.
%
%
%
%
%
%
      \begin{table}
        \centering
        \fontsize{8}{11pt}\selectfont
        \tabcolsep 0.11truecm
        \caption{Observational details of the visible spectra (1800~s each) obtained for (474640)~2004~VN$_{112}$ and 2013~RF$_{98}$ (see 
                 the text for further details).
                }
        \begin{tabular}{cccccc}
          \hline
             Target          & Spec \# &   Date   & UT Start & Airmass & Slope (\%/0.1~$\mu$m)  \\
          \hline
                             & 1       & 09-03-16 &  04:37   &  1.085  &      10.2$\pm$0.6      \\
             474640          & 2       & 09-03-16 &  05:07   &  1.072  &      13.3$\pm$0.6      \\
                             &         &          &          &\bf{Mean}& \bf{12$\mathbf{\pm}$2} \\
          \hline
                             & 1       & 09-08-16 &  03:07   &  1.234  &      14.9$\pm$0.8      \\
                             & 2       & 09-08-16 &  03:37   &  1.181  &      12.3$\pm$0.8      \\              
             2013~RF$_{98}$  & 3       & 09-08-16 &  04:07   &  1.151  &      17.1$\pm$0.8      \\
                             & 4       & 09-08-16 &  04:38   &  1.142  &      16.1$\pm$0.8      \\
                             &         &          &          &\bf{Mean}& \bf{15$\mathbf{\pm}$2} \\
          \hline
        \end{tabular}
        \label{spec}
      \end{table}
%
%

\begin{figure}
        \centering
         \includegraphics[width=\linewidth]{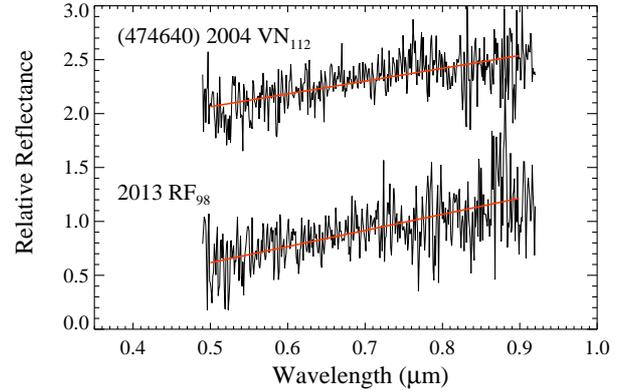}
         \caption{Final visible spectra of the pair of ETNOs (see the text for details). 
                 }
         \label{specBOTH}
      \end{figure}

        In Table \ref{spec}, the spectral slope (in units of \%/0.1~$\mu$m) has been computed from a linear fitting to the spectrum in the 
        wavelength range 0.5--0.9 $\mu$m. We used an iterative process ---removing a total of 50 points randomly distributed in the spectrum 
        and performing a linear fitting--- and obtained a value of the slope for each iteration. The resulting slope value is the mean of a 
        total of 100 iterations and the associated error is the standard deviation of this mean. The process of dividing by the spectra of 
        the solar analogue stars introduces an error of 0.3~\%/0.1~$\mu$m. The error value shown in the last column of Table \ref{spec} is 
        sum of these two contributions. For each target, we averaged individual spectra to obtain the average visible spectra for 474640 and 
        2013~RF$_{98}$ shown in Fig. \ref{specBOTH}; the mean spectral slopes in Table \ref{spec} are the means of the individual spectra 
        and their errors, the associated standard deviations. The value of the spectral slope of 474640 is consistent with the one obtained 
        by Sheppard (2010). The values in Table \ref{spec} are similar to those of scattered disc TNOs, Plutinos, high-inclination classical 
        TNOs as well as the Damocloids and comets (see e.g. table 5 in Sheppard 2010).
%
%
%
%

\begin{figure}
        \centering
         \includegraphics[width=\linewidth]{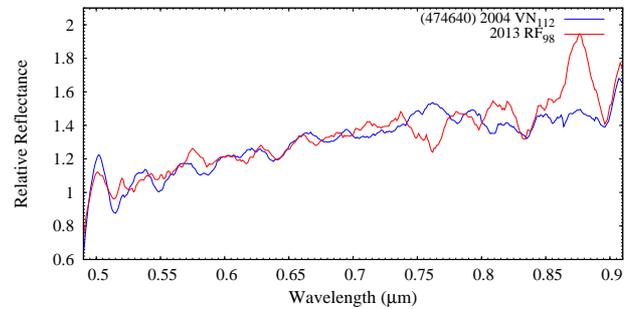}
         \caption{Comparison between the spectra of (474640)~2004~VN$_{112}$ and 2013~RF$_{98}$ smoothed by a Savitzky-Golay filter (see the 
                  text) and scaled to match at 0.60~$\mu$m. The most prominent absorption band of pure methane ice at 0.73~$\mu$m is not 
                  seen on either spectra. 
                 }
         \label{savgol}
      \end{figure}

        The spectra in Fig. \ref{specBOTH} show an obvious resemblance, but the low S/N makes the identification of absorption features 
        difficult. In order to make a better comparison and assuming that the relative reflectance is both slowly varying and corrupted by 
        random noise, we applied a Savitzky-Golay filter (Savitzky \& Golay 1964) to both datasets. Such filters provide smoothing without 
        loss of resolution, approximating the underlying function by a polynomial as described by e.g. Press et al. (1992). Fig. \ref{savgol} 
        shows the smoothed spectra after applying a 65 point Savitzky-Golay filter of order 6 to the data in Fig. \ref{specBOTH}. The 
        smoothed spectra, in particular that of 474640 (blue), show some weak features that might be tentatively identified as pure methane 
        ice absorption bands (Grundy, Schmitt \& Quirico 2002). However, the most prominent methane band at 0.73~$\mu$m is observed neither 
        in the spectrum of 474640 nor in that of 2013~RF$_{98}$ (red). The S/N is insufficient to identify reliably any absorption band, but 
        the spectral slopes in the visible of both objects provide some compositional information. Objects with visible spectral slopes in 
        the range 0--10~\%/0.1~$\mu$m can have pure ices on their surfaces (like Eris, Pluto, Makemake and Haumea), as well as highly 
        processed carbon. Slightly red slopes (5--15~\%/0.1~$\mu$m) indicate the possible presence of amorphous silicates as in the case of 
        Trojans (Emery \& Brown 2004) or Thereus (Licandro \& Pinilla-Alonso 2005). In any case, the objects will not have a surface 
        dominated by complex organics (tholins). Differences between the spectra of 474640 and 2013~RF$_{98}$ might be the result of their 
        present-day heliocentric distance. Mechanisms that are more efficient in altering the icy surfaces of these objects at smaller 
        perihelion distances include sublimation of volatiles and micrometeoroid bombardment (Santos-Sanz et al. 2009).
%
%
%
%

     \subsection{Photometry}
        We obtained a total of 3 and 11 acquisition images to identify (474640)~2004~VN$_{112}$ and 2013~RF$_{98}$, respectively. Images 
        were taken using the Sloan $r'$ filter and calibrated using the zero point values computed for the corresponding nights. The 
        resulting $r'$ magnitudes and their uncertainties are shown in Table \ref{phot}.

     \subsection{Astrometry}
        We used the acquisition images to compute the celestial coordinates of each target and improve its orbit. This was particularly 
        relevant for 2013~RF$_{98}$ that prior to our observations had a rather uncertain orbital determination.$^{2}$ We found 
        (474640)~2004~VN$_{112}$ within 1\arcsec of the predicted ephemerides, but 2013~RF$_{98}$ was found nearly 1\farcm2 away. Both ETNOs 
        were in Cetus. Astrometric calibration of the CCD frames was performed using the algorithms of the {\it Astrometry.net} system (Lang 
        et al. 2010). The quality of high-precision astrometry with OSIRIS at GTC matches that of data acquired with FORS2/VLT (Sahlmann et 
        al. 2016). The collected astrometry is shown in Tables \ref{astrometry2004VN112} and \ref{astrometry2013RF98}. The new orbital 
        solution for 2013~RF$_{98}$ available from \textsc{Horizons} (as of 2016 December 18 18:51:59 UT) is based on 51 observations 
        spanning a data-arc of 1092 d, its residual rms amounts to 0\farcs12: $a=349\pm11$ au, $e=0.897\pm0.003$, $i=29\fdg572\pm0\fdg003$, 
        $\Omega=67\fdg596\pm0\fdg005$ and $\omega=311\fdg8\pm0\fdg6$, referred to the epoch 2457800.5 JD TDB; the time of perihelion passage 
        is 2455125$\pm$95 JED (2009 October 20.7289 UT). The time of perihelion passage for 474640 is 2009 August 25.8290 UT. Using the 
        new orbit, the directions of the perihelia of this pair are separated by 14\fdg2$\pm$0\fdg6, their orbital poles by 
        4\fdg056$\pm$0\fdg003, and the directions of their velocities at perihelion/aphelion by 14\fdg1$\pm$0\fdg6. 

  \section{Origin of the pair 474640--2013~RF$_\mathbf{98}$: fragmentation vs. binary dissociation}
     From the spectral analysis discussed above, we have found that the members of the pair (474640)~2004~VN$_{112}$--2013~RF$_{98}$ show 
     similar spectral slopes, very different from that of Sedna which has ultra-red surface material (spectral gradient of about 
     26~\%/0.1~$\mu$m according to Sheppard 2010, and 42~\%/0.1~$\mu$m according to Fornasier et al. 2009) but compatible with those of 
     (148209)~2000~CR$_{105}$ (spectral gradient of about 14~\%/0.1~$\mu$m, Sheppard 2010) and 2012~VN$_{113}$ (spectral gradient of about 
     13~\%/0.1~$\mu$m, Trujillo \& Sheppard 2014). These five objects have been included in the group of seven singled out as relevant to 
     the Planet Nine hypothesis (Brown \& Batygin 2016). Such spectral differences suggest that the region of origin of the pair 
     474640--2013~RF$_{98}$ may coincide with that of 148209 and 2012~VN$_{113}$ but not with Sedna's, which is thought to come from the 
     inner Oort Cloud (Sheppard 2010). Other ETNOs with values of their spectral gradient in Sheppard (2010) are 2002~GB$_{32}$ 
     ($\sim$17~\%/0.1~$\mu$m) and 2003~HB$_{57}$ ($\sim$13~\%/0.1~$\mu$m).

     Objects with both similar directions of the orbital poles and perihelia could be part of a group of common physical origin (\"Opik 
     1971). This particular pair of ETNOs is very unusual and a model analogous to the one used by de la Fuente Marcos \& de la Fuente 
     Marcos (2014) to study the overall visibility of the ETNO population predicts that the probability of finding such a pair by chance is 
     less than 0.0002. This model uses the new orbital solutions and assumes an unperturbed asteroid population moving in heliocentric 
     orbits. Following \"Opik (1971), there are two independent scenarios that could explain this level of coincidence: (1) a large object 
     broke up relatively recently at perihelion and these two ETNOs are fragments, or (2) both ETNOs were kicked by an unseen perturber at 
     aphelion. Sekanina (2001) has shown that minor bodies resulting from a fragmentation episode at perihelion must have very different 
     times of perihelion passage. The fragmentation episode at perihelion can thus be readily discarded as the difference in time of 
     perihelion passage for this pair is less than a year. 

     The second scenario pointed out above implies the presence of an unseen massive perturber, i.e. a trans-Plutonian planet. Close 
     encounters between minor bodies and planets can induce fragmentation directly via tidal forces (e.g. Sharma, Jenkins \& Burns 2006) or 
     indirectly by exciting rapid rotation (e.g. Scheeres et al. 2000; Ortiz et al. 2012). Alternatively, wide binary asteroids can be 
     easily disrupted during close encounters with planets. The existence of wide binaries among the populations of minor bodies orbiting 
     beyond Neptune is well documented (e.g. Parker et al. 2011). Wide binary asteroids have very low binding energies and can be easily 
     dissociated during close encounters with planets (e.g. Agnor \& Hamilton 2006; Parker \& Kavelaars 2010). Binary asteroid dissociation 
     may be able to explain the properties of this pair of ETNOs, but only if there is a massive unseen perturber orbiting the Sun well 
     beyond Pluto.
%
%
      \begin{figure}
        \centering
         \includegraphics[width=\linewidth]{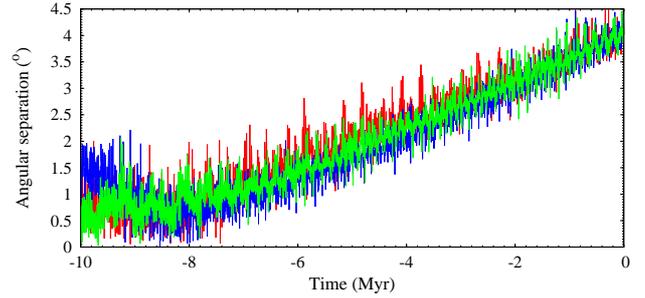}
         \caption{Evolution of the angular separation between the orbital poles of the pair (474640)~2004~VN$_{112}$--2013~RF$_{98}$ for three 
                  representative test calculations with different perturbers. Red: $a$ = 549 au, $e$ = 0.21, $i$ = 47\degr, $m$ = 16 
                  $M_{\oplus}$. Blue: $a$ = 448 au, $e$ = 0.16, $i$ = 33\degr, $m$ = 12 $M_{\oplus}$. Green: $a$ = 421 au, $e$ = 0.10, 
                  $i$ = 33\degr, $m$ = 12 $M_{\oplus}$.
                 }
         \label{seppoles}
      \end{figure}
%
%

     In order to test the viability of this hypothesis, we have performed thousands of numerical experiments following the prescriptions 
     discussed by de la Fuente Marcos et al. (2016) and aimed at finding the most probable orbital properties of a putative perturber able 
     to tilt the orbital plane of the pair 474640--2013~RF$_{98}$ from an initial angular separation close to zero at dissociation to the 
     current value of nearly 4\degr. These simulations involve $N$-body integrations backwards in time under the influence of an unseen 
     perturber with varying orbital and physical parameters (per numerical experiment). Our preliminary results indicate that a planet with 
     mass, $m$, in the range 10--20 $M_{\oplus}$ moving in an eccentric (0.1--0.4) and inclined (20--50\degr) orbit with semimajor axis of 
     300--600 au, may be able to induce the observed tilt on a time-scale of 5--10 Myr. Perturbers with $m<10\ M_{\oplus}$ or $a>600$ au are 
     unable to produce the desired effect. Fig. \ref{seppoles} illustrates the typical outcome of these calculations. A detailed account of 
     these numerical experiments will be presented in an accompanying paper (de la Fuente Marcos, de la Fuente Marcos \& Aarseth, in 
     preparation). The orbital parameters of this putative planet are somewhat consistent with those of the object discussed by Holman \& 
     Payne (2016). Super-Earths may form at 125--750 au from the Sun (Kenyon \& Bromley 2015, 2016). Our analysis favours a scenario in 
     which 474640--2013~RF$_{98}$ were once a binary asteroid that became unbound after a relatively recent gravitational encounter with a 
     trans-Plutonian planet at hundreds of au from the Sun. An alternative explanation involving an asteroid break-up near aphelion, also 
     after a close encounter with a planet, is possible but less probable because it requires an approach at closer range, 20 planetary 
     radii versus 0.8~au for binary dissociation.

  \section{Conclusions}
     In this Letter, we provide for the first time direct indication of the surface composition of the ETNOs (474640)~2004~VN$_{112}$ and 
     2013~RF$_{98}$. Both objects are too faint for infrared spectroscopy, but our results show that they are viable targets for visible 
     spectroscopy. The analysis of our results gives further support to the trans-Plutonian planets paradigm that predicts the presence of 
     one or more planetary bodies well beyond Pluto. Summarizing: 
     \begin{enumerate}[(i)]
        \item Our estimate of the spectral slope for 474640 is 12$\pm$2~\%/0.1~$\mu$m and for 2013~RF$_{98}$ is 15$\pm$2~\%/0.1~$\mu$m. 
              These values suggest that the surfaces of these ETNOs can have pure methane ices (like Pluto) and highly processed carbons, 
              including some amorphous silicates.
        \item Although the spectra of the pair 474640--2013~RF$_{98}$ are not perfect matches, the resemblance is significant and
              the disparities observed might be the result of their different present-day heliocentric distance.
        \item By improving the orbital solution of 2013~RF$_{98}$, we confirm that the pair 474640--2013~RF$_{98}$ has unusual 
              relative dynamical properties. The directions of their perihelia are separated by 14\fdg2 and their orbital poles are 4\fdg1 
              apart. 
        \item Our numerical analysis favours a scenario in which 474640--2013~RF$_{98}$ were once a binary asteroid that became 
              unbound after an encounter with a trans-Plutonian planet at very large heliocentric distance.  
     \end{enumerate}

  \section*{Acknowledgements}
     We thank the anonymous referee for a constructive and detailed report, and S.~J. Aarseth for providing one of the codes used in this 
     research and for his helpful comments on binary dissociation. Based on observations made with the Gran Telescopio Canarias; we are 
     grateful to all the technical staff and telescope operators for their assistance with the observations. This work was partially 
     supported by the Spanish `Ministerio de Econom\'{\i}a y Competitividad' (MINECO) under grant ESP2014-54243-R. JdL acknowledges 
     financial support from MINECO under the 2015 Severo Ochoa Program MINECO SEV-2015-0548. CdlFM and RdlFM thank A. I. G\'omez de Castro, 
     I. Lizasoain and L. Hern\'andez Y\'a\~nez of the Universidad Complutense de Madrid (UCM) for providing access to computing facilities. 
     Part of the calculations and the data analysis were completed on the EOLO cluster of the UCM, and CdlFM and RdlFM thank S. Cano Als\'ua 
     for his help during this stage. EOLO, the HPC of Climate Change of the International Campus of Excellence of Moncloa, is funded by the 
     MECD and MICINN. This is a contribution to the CEI Moncloa. In preparation of this Letter, we made use of the NASA Astrophysics Data 
     System, the ASTRO-PH e-print server and the MPC data server.

  \newpage
  \appendix
  \section{Photometry and astrometry data tables}
     The $r'$ magnitudes from the acquisition images and their uncertainties are shown in Table \ref{phot}. The astrometry submitted 
     to the Minor Planet Center (MPC) is shown in Tables \ref{astrometry2004VN112} and \ref{astrometry2013RF98} (de Le\'on, de la Fuente Marcos
     \& de la Fuente Marcos 2016).
%
%
      \begin{table}
        \centering
        \fontsize{8}{11pt}\selectfont
        \tabcolsep 0.08truecm
        \caption{Observational details of the acquisition images obtained to identify the targets.
                }
        \begin{tabular}{ccccccc}
          \hline
             Image \#        &   Date   & UT Start & Airmass & Exp. Time & Zero point &     $r'$        \\
                             &          &          &         &    (s)    &   (mag)    &     (mag)       \\
          \hline
                             &          &          &         &           &            & (474640)        \\
                             &          &          &         &           &            & 2004~VN$_{112}$ \\
          \hline
             1               & 09-02-16 &   02:16  &  1.580  &    120    &  29.240    & 23.63$\pm$0.10  \\
             2               & 09-03-16 &   04:21  &  1.120  &    120    &  29.277    & 23.58$\pm$0.05  \\
             3               & 09-03-16 &   04:32  &  1.100  &     60    &  29.277    & 23.51$\pm$0.05  \\
          \hline
                             &          &          &         &           &            & 2013~RF$_{98}$  \\
          \hline
             1               & 09-05-16 &   02:43  &  1.380  &    180    &  29.281    & 24.57$\pm$0.10  \\
             2               & 09-06-16 &   02:55  &  1.320  &    120    &  29.287    & 24.39$\pm$0.20  \\
             3               & 09-06-16 &   03:05  &  1.290  &    120    &  29.287    & 24.42$\pm$0.15  \\
             4               & 09-06-16 &   03:23  &  1.240  &    120    &  29.287    & 24.51$\pm$0.20  \\
             5               & 09-06-16 &   03:32  &  1.220  &    120    &  29.287    & 24.45$\pm$0.05  \\
             6               & 09-06-16 &   03:37  &  1.210  &    120    &  29.287    & 24.53$\pm$0.20  \\
             7               & 09-07-16 &   03:40  &  1.210  &    180    &  29.287    & 24.53$\pm$0.05  \\
             8               & 09-08-16 &   02:46  &  1.330  &    120    &  29.287    & 24.56$\pm$0.07  \\
             9               & 09-08-16 &   02:51  &  1.310  &    180    &  29.287    & 24.55$\pm$0.10  \\
            10               & 09-08-16 &   02:56  &  1.290  &    120    &  29.287    & 24.53$\pm$0.05  \\
            11               & 09-08-16 &   03:00  &  1.280  &    120    &  29.287    & 24.49$\pm$0.10  \\
          \hline
        \end{tabular}
        \label{phot}
      \end{table}
%
%
%
%
     \begin{table}
        \fontsize{8}{11pt}\selectfont
        \tabcolsep 0.20truecm
        \caption{Observations of (474640)~2004~VN$_{112}$. All the observations in the $r'$ filter.} 
        \centering
        \begin{tabular}{ccc}
           \hline
            Date             & RA(J2000)                          & Dec(J2000)              \\
            (UT)             & ($^{\rm h}$:$^{\rm m}$:$^{\rm s}$) & (\degr:\arcmin:\arcsec) \\
           \hline
            2016 09 02.09514 & 03:07:52.51                        & +07:38:28.1             \\
            2016 09 03.18233 & 03:07:51.41                        & +07:38:17.7             \\
            2016 09 03.18923 & 03:07:51.40                        & +07:38:17.5             \\
           \hline
        \end{tabular}
        \label{astrometry2004VN112}
     \end{table}
%
%
%
%
     \begin{table}
        \fontsize{8}{11pt}\selectfont
        \tabcolsep 0.20truecm
        \caption{Observations of 2013~RF$_{98}$. All the observations in the $r'$ filter.} 
        \centering
        \begin{tabular}{ccc}
           \hline
            Date              & RA(J2000)                          & Dec(J2000)              \\
            (UT)              & ($^{\rm h}$:$^{\rm m}$:$^{\rm s}$) & (\degr:\arcmin:\arcsec) \\
           \hline
            2016 09 05.114385 & 02:49:09.83                        & $-$00:10:52.0           \\
            2016 09 06.122606 & 02:49:07.83                        & $-$00:11:14.2           \\  
            2016 09 06.129458 & 02:49:07.81                        & $-$00:11:14.5           \\
            2016 09 06.142113 & 02:49:07.79                        & $-$00:11:14.7           \\
            2016 09 06.148191 & 02:49:07.78                        & $-$00:11:14.7           \\
            2016 09 06.151823 & 02:49:07.77                        & $-$00:11:14.8           \\
            2016 09 07.154454 & 02:49:05.70                        & $-$00:11:37.3           \\
            2016 09 08.116547 & 02:49:03.61                        & $-$00:11:59.1           \\
            2016 09 08.120114 & 02:49:03.60                        & $-$00:11:59.1           \\
            2016 09 08.123483 & 02:49:03.60                        & $-$00:11:59.3           \\
            2016 09 08.126326 & 02:49:03.59                        & $-$00:11:59.4           \\
           \hline
        \end{tabular}
        \label{astrometry2013RF98}
     \end{table}
%
%

  \bsp
  \label{lastpage}

\begin{thebibliography}{99}
     \bibitem[\protect\citeauthoryear{Abbott et al.}{2005}]{AB05} Abbott T. et al., 2005,
             The Dark Energy Survey,
             eprint arXiv:astro-ph/0510346
     \bibitem[\protect\citeauthoryear{Abbott et al.}{2016}]{AB16} Abbott T. et al., 2016,
             MNRAS, 460, 1270
     \bibitem[\protect\citeauthoryear{Agnor \& Hamilton}{2006}]{AH06} Agnor C. B., Hamilton D. P., 2006,
             Nature, 441, 192
     \bibitem[\protect\citeauthoryear{Batygin \& Brown}{2016}]{BB16} Batygin K., Brown M. E., 2016,
             AJ, 151, 22
     \bibitem[\protect\citeauthoryear{Becker et al.}{2007}]{BE07} Becker A. C., Arraki K. S., Rest A., Wood-Vasey W. M., Marsden B. G., 2007,
             MPEC Circ., MPEC 2007-S29
     \bibitem[\protect\citeauthoryear{Becker et al.}{2008}]{BE08} Becker A. C. et al., 2008,
             ApJ, 682, L53
     \bibitem[\protect\citeauthoryear{Brown}{2012}]{BR12} Brown M. E., 2012,
             Annu. Rev. Earth Planet. Sci., 40, 467  
     \bibitem[\protect\citeauthoryear{Brown \& Batygin}{2016}]{BR16} Brown M. E., Batygin K., 2016,
             ApJ, 824, L23 
     \bibitem[\protect\citeauthoryear{Cepa et al.}{2000}]{CE00} Cepa J. et al., 2000,
             in Iye M., Moorwood A. F. M., eds, Optical and IR Telescope Instrumentation and Detectors.
             Proc. SPIE, 4008, p.\ 623
     \bibitem[\protect\citeauthoryear{de la Fuente Marcos \& de la Fuente Marcos}{2014}]{FM14} de la Fuente Marcos C., de la Fuente Marcos R., 2014,
             MNRAS, 443, L59
     \bibitem[\protect\citeauthoryear{de la Fuente Marcos \& de la Fuente Marcos}{2016}]{FM16} de la Fuente Marcos C., de la Fuente Marcos R., 2016,
             MNRAS, 462, 1972
     \bibitem[\protect\citeauthoryear{de la Fuente Marcos et al.}{2015}]{MA15} de la Fuente Marcos C., de la Fuente Marcos R., Aarseth S. J., 2015,
             MNRAS, 446, 1867
     \bibitem[\protect\citeauthoryear{de la Fuente Marcos et al.}{2016}]{MA16} de la Fuente Marcos C., de la Fuente Marcos R., Aarseth S. J., 2016,
             MNRAS, 460, L123
     \bibitem[\protect\citeauthoryear{de Le\'on et al.}{2016}]{LC16} de Le\'on J. et al., 2016,
             Icarus, 266, 57
     \bibitem[\protect\citeauthoryear{de Le\'on, de la Fuente Marcos \& de la Fuente Marcos}{2016}]{LM16} de Le\'on J., de la Fuente Marcos C., 
             de la Fuente Marcos R., 2016,
             MPEC Circ., MPEC 2016-U18
     \bibitem[\protect\citeauthoryear{Emery \& Brown}{2004}]{EB04} Emery J. P., Brown R. H., 2004, 
             Icarus, 170, 131
     \bibitem[\protect\citeauthoryear{Flaugher et al.}{2015}]{FL15} Flaugher B. et al., 2015,
             AJ, 150, 150
     \bibitem[\protect\citeauthoryear{Fornasier et al.}{2009}]{FO09} Fornasier S. et al., 2009,
             A\&A, 508, 457
     \bibitem[\protect\citeauthoryear{Fraser \& Brown}{2012}]{FB12} Fraser W. C., Brown M. E., 2012,
             ApJ, 749, 33
     \bibitem[\protect\citeauthoryear{Gladman et al.}{2002}]{GL02} Gladman B., Holman M., Grav T., Kavelaars J., Nicholson P., 
             Aksnes K., Petit J.-M., 2002, 
             Icarus, 157, 269
     \bibitem[\protect\citeauthoryear{Grundy, Schmitt \& Quirico}{2002}]{GR02} Grundy W., Schmitt B., Quirico E., 2002, 
             Icarus, 144, 486
     \bibitem[\protect\citeauthoryear{Holman \& Payne}{2016}]{HP16} Holman M. J., Payne M. J., 2016,
             AJ, 152, 80
     \bibitem[\protect\citeauthoryear{Kenyon \& Bromley}{2015}]{KB15} Kenyon S. J., Bromley B. C., 2015,
             ApJ, 806, 42
     \bibitem[\protect\citeauthoryear{Kenyon \& Bromley}{2016}]{KB16} Kenyon S. J., Bromley B. C., 2016,
             ApJ, 825, 33 
     \bibitem[\protect\citeauthoryear{Landolt}{1992}]{LA92} Landolt A. U., 1992,
             AJ, 104, 340 
     \bibitem[\protect\citeauthoryear{Lang et al.}{2010}]{LA10} Lang D., Hogg D. W., Mierle K., Blanton M., Roweis S., 2010,
             AJ, 139, 1782
     \bibitem[\protect\citeauthoryear{Licandro \& Pinilla-Alonso}{2005}]{LP05} Licandro J., Pinilla-Alonso N., 2005, 
             ApJ, 630, L93
     \bibitem[\protect\citeauthoryear{Morate et al.}{2016}]{MO16} Morate D., de Le\'on J., De Pr\'a M., Licandro J., 
             Cabrera-Lavers A., Campins H., Pinilla-Alonso N., Al\'{\i}-Lagoa V., 2016,
             A\&A, 586, A129
     \bibitem[\protect\citeauthoryear{Morbidelli \& Levison}{2004}]{ML04} Morbidelli A., Levison H. F., 2004,
             AJ, 128, 2564
     \bibitem[\protect\citeauthoryear{\"Opik}{1971}]{OP71} \"Opik E. J., 1971,
             Irish Astronomical Journal, 10, 35
     \bibitem[\protect\citeauthoryear{Ortiz et al.}{2012}]{OR12} Ortiz J. L. et al., 2012,
             MNRAS, 419, 2315
     \bibitem[\protect\citeauthoryear{Parker \& Kavelaars}{2010}]{PK10} Parker A. H., Kavelaars J. J., 2010,
             ApJ, 722, L204
     \bibitem[\protect\citeauthoryear{Parker et al.}{2011}]{PA11} Parker A. H., Kavelaars J. J., Petit J.-M., Jones L., Gladman B., Parker J., 2011,
             ApJ, 743, 1
     \bibitem[\protect\citeauthoryear{Press et al.}{1992}]{PR92} Press W. H., Teukolsky S. A., Vetterling W. T., Flannery B. P., 1992,
             Numerical Recipes in FORTRAN.
             Cambridge Univ. Press, Cambridge, p.\ 644
     \bibitem[\protect\citeauthoryear{Sahlmann et al.}{2016}]{SA16} Sahlmann J., Lazorenko P. F., Bouy H., Mart\'{\i}n E. L., Queloz D.,
             S\'egransan D., Zapatero Osorio M. R., 2016,
             MNRAS, 455, 357
     \bibitem[\protect\citeauthoryear{Santos-Sanz et al.}{2009}]{SS09} Santos-Sanz P., Ortiz J. L., Barrera L., Boehnhardt H., 2009,
             A\&A, 494, 693 
     \bibitem[\protect\citeauthoryear{Savitzky \& Golay}{1964}]{SG64} Savitzky A., Golay M. J. E., 1964,
             Analytical Chemistry, 36, 1627
     \bibitem[\protect\citeauthoryear{Scheeres et al.}{2000}]{SC00} Scheeres D. J., Ostro S. J., Werner R. A., Asphaug E., Hudson R. S., 2000,
             Icarus, 147, 106
     \bibitem[\protect\citeauthoryear{Sekanina}{2001}]{SE01} Sekanina Z., 2001,
             Publications of the Astronomical Institute of the Academy of Sciences of the Czech Republic, 89, 78
     \bibitem[\protect\citeauthoryear{Shankman et al.}{2016}]{SH16} Shankman C., Kavelaars J. J., Lawler S. M., Gladman B. J., Bannister M. T., 2016,
             AJ, submitted (eprint arXiv:1610.04251)
     \bibitem[\protect\citeauthoryear{Sharma et al.}{2006}]{SH06} Sharma I., Jenkins J. T., Burns J. A., 2006,
             Icarus, 183, 312
     \bibitem[\protect\citeauthoryear{Sheppard}{2010}]{SH10} Sheppard S. S., 2010,
             AJ, 139, 1394
     \bibitem[\protect\citeauthoryear{Sheppard \& Trujillo}{2016}]{ST16} Sheppard S. S., Trujillo C. A., 2016,
             AJ, 152, 221
     \bibitem[\protect\citeauthoryear{Trujillo \& Sheppard}{2014}]{TS14} Trujillo C. A., Sheppard S. S., 2014,
             Nature, 507, 471
  \end{thebibliography}
\end{document}